\documentclass[12pt]{article}
\usepackage{epsfig}
\usepackage{psfrag}
\usepackage{latexsym}
\def\be{\begin{equation}}
\def\ee{\end{equation}}
\def\bc{\begin{center}}
\def\ec{\end{center}}
\def\bea{\begin{eqnarray}}
\def\eea{\end{eqnarray}}
\def\dd{\displaystyle}
\def\nn{\nonumber}

\catcode`@=11
\def\marginnote#1{}
\newcount\hour
\newcount\minute
\newtoks\amorpm
\hour=\time\divide\hour by60
\minute=\time{\multiply\hour by60 \global\advance\minute by-\hour}
\edef\standardtime{{\ifnum\hour<12 \global\amorpm={am}%
        \else\global\amorpm={pm}\advance\hour by-12 \fi
        \ifnum\hour=0 \hour=12 \fi
        \number\hour:\ifnum\minute<10 0\fi\number\minute\the\amorpm}}
\edef\militarytime{\number\hour:\ifnum\minute<10 0\fi\number\minute}
\def\draftlabel#1{{\@bsphack\if@filesw {\let\thepage\relax
   \xdef\@gtempa{\write\@auxout{\string
      \newlabel{#1}{{\@currentlabel}{\thepage}}}}}\@gtempa
   \if@nobreak \ifvmode\nobreak\fi\fi\fi\@esphack}
        \gdef\@eqnlabel{#1}}
\def\@eqnlabel{}
\def\@vacuum{}
\def\draftmarginnote#1{\marginpar{\raggedright\scriptsize\tt#1}}
\def\draft{\oddsidemargin 0.0truein
        \def\@oddfoot{\sl preliminary draft \hfil
        \rm\thepage\hfil\sl\today\quad\militarytime}
        \let\@evenfoot\@oddfoot \overfullrule 3pt
        \let\label=\draftlabel
        \let\marginnote=\draftmarginnote
   \def\@eqnnum{(\theequation)\rlap{\kern\marginparsep\tt\@eqnlabel}%
\global\let\@eqnlabel\@vacuum}  }
\catcode`@=12
\begin{document}
\begin{titlepage}
\vspace*{-1cm}
\phantom{hep-ph/0305129} 
\hfill{DFPD-03/TH/16}

\hfill{SACLAY-T03/059}

\hfill{UG-FT-150/03}

\hfill{CAFPE-20-03}
\vskip 0.5cm
\begin{center}
{\Large\bf Fermion Generations, Masses and Mixing Angles\\
\vskip .1cm
from Extra Dimensions}
\end{center}
\vskip 0.5  cm
\begin{center}
{\large Carla Biggio}~\footnote{e-mail address: biggio@pd.infn.it}
~~~{\large Ferruccio Feruglio}~\footnote{e-mail address: feruglio@pd.infn.it}
\\
\vskip .1cm
Dipartimento di Fisica `G.~Galilei', Universit\`a di Padova 
\\ 
INFN, Sezione di Padova, Via Marzolo~8, I-35131 Padua, Italy
\\
\vskip .2cm
{\large Isabella Masina}~\footnote{e-mail address: masina@spht.saclay.cea.fr}$^,$ \footnote{address after May 2003: Enrico Fermi Center,
Via Panisperna 89/A 00184 Roma, Italy}
\\
\vskip .1cm
Service de Physique Th\'eorique CEA/Saclay\\
Orme des Merisiers 
F-91191 Gif-sur-Yvette Cedex, France 
\\
\vskip .2cm
{\large Manuel P\'erez-Victoria}~\footnote{e-mail address: manolo@pd.infn.it}
\\
\vskip .1cm
Dipartimento di Fisica `G.~Galilei', Universit\`a di Padova 
\\ 
INFN, Sezione di Padova, Via Marzolo~8, I-35131 Padua, Italy
\\
Departamento de F\'{\i}sica Te\'orica y del Cosmos\\
Centro Andaluz de F\'{\i}sica de Part\'{\i}culas Elementales (CAFPE)\\
Universidad de Granada, E-18071 Granada, Spain
\end{center}
\vskip 0.7cm
\begin{abstract}
\noindent
We discuss a toy model in six dimensions that predicts 
two fermion generations, natural mass hierarchy
and intergenerational mixing. Matter is described
by vector-like six dimensional fermions, one per each
irreducible standard model representation. Two fermion
generations arise from the compactification mechanism,
through orbifold projection. They are localized in
different regions of the compact space by a six
dimensional mass term. Flavour symmetry is broken via
Yukawa couplings, with a Higgs vacuum expectation
value not constant in the extra space. A hierarchical
spectrum is obtained from order one dimensionless
parameters of the six dimensional theory. The Cabibbo angle
arises from the soft breaking of six dimensional parity
symmetry. We also briefly discuss how the present model
could be extended to cover the realistic case.
\end{abstract}
\end{titlepage}
\setcounter{footnote}{0}
\vskip2truecm

%
%
\bc
{\bf 1. Introduction}
\ec
In the last years our knowledge in flavour in physics
has undergone a spectacular development. On the experimental side, 
with the data from the existing B factories, BABAR and Belle, we entered an era
of precision tests in the quark sector. 
For instance the $\vert V_{cb}\vert$ element of the quark mixing matrix 
is now known with a precision of few percents \cite{pdg}. Moreover many
independent measurements are now over-constraining the quark parameters
of the standard model (SM), and the success of the theory 
in fitting all the data is really impressive.
Also the picture in the lepton sector has been
greatly clarified and, even if we have not yet obtained precise
determination of mass and mixing parameters, nevertheless
a clear pattern has been identified from the solutions to
the solar and atmospheric neutrino problems \cite{fog}.

On the theoretical side, we should honestly admit that
flavour still represents one of the great mysteries
in particle physics. We do not know the scale at which
the flavour dynamics sets in. 
Perhaps at this scale
a conventional, four dimensional, picture still holds
thus allowing us to analyze the flavour problem in the context
of a local quantum field theory in four space-time dimensions.
Here the most powerful tool that we have to decipher
the observed hierarchy among the different masses and mixing angles
is that of spontaneously broken flavour symmetries \cite{fn}.
In the idealized limit of exact symmetry, only the heaviest
fermions are massive: the top quark and, maybe, the whole
third family. The lightest fermions and the small mixing angles
originate from breaking effects. This beautiful idea has
been widely explored in many possible versions, with discrete
or continuous symmetries, global or local ones. 
A realistic description of fermion masses in this framework
typically requires either a large number of parameters
or a high degree of complexity and we are probably unable to 
select the best model among the many existing ones. Moreover,
in four dimensions we have little hopes to understand why
there are exactly three generations. 

It might be the case that at the energy scale characterizing
flavour physics a four-dimensional description breaks down.
For instance this happens in superstring theories where 
the space-time is ten or eleven dimensional.
In the ten dimensional heterotic string six dimensions
can be compactified on a Calabi-Yau manifold \cite{can} or on orbifolds
\cite{dix} and the flavour properties are strictly related to the
features of the compact space. In Calabi-Yau compactifications
the number of chiral generations is proportional to the Euler
characteristics of the manifold. In orbifold compactifications,
matter in the twisted sector is localized around the orbifold
fixed points and the Yukawa couplings, arising from world-sheet
instantons, have a natural geometrical interpretation \cite{orb}.
Recently string realizations where the light matter fields 
of the standard model
arises from intersecting branes have been proposed. 
Also in this context the flavour dynamics is controlled by
topological properties of the geometrical construction \cite{int}, having no 
counterpart in four dimensional field theories.

Perhaps in the future the flavour mystery will be unraveled by string theory,
but in the meantime it would be interesting to explore, 
in a pure field theoretical construction,
the possibility of extra space-like dimensions.
We can then take advantage of the greater freedom that a bottom-up
field-theory approach possesses compared to string theory. 
Moreover in the last years a lot of progress has been done
in understanding field theories with extra spatial dimensions. 
These theories are ultraviolet divergent and should be cut-off
at some energy scale $\Lambda$, but they can still be useful
as effective descriptions at low energies, including the 
compactification scale.
Semi-realistic models have been proposed within orbifold compactification,
allowing for light chiral fermions \cite{orbmod,orb6d}.
The compactification mechanism and 
the orbifold projection have 
also been exploited to break supersymmetry \cite{susyb,susybm} 
and/or gauge symmetry
\cite{gaugeb,gaugebm} with distinctive and attractive features \cite{ftorb}.

It has soon been realized that also in a field theoretical description
the existence of extra dimensions could have important consequences
for the flavour problem. For instance in orbifold compactifications
light four dimensional fermions may be either localized at the orbifold
fixed points or they may arise as zero modes of higher-dimensional 
spinors, with a wave function suppressed by the square root of the 
volume of the compact space. This led to several interesting proposals.
It has been suggested that the smallness of neutrino masses could
be reproduced if the left-handed
active neutrinos sit at a fixed point and the right-handed
sterile partners live in the bulk of a large fifth dimension \cite{nuex}.
In five dimensional grand unified theories the heaviness of the
third generation can be explained by localizing the corresponding
fields on a fixed point, whereas the relative lightness of the
first two generations as well as the breaking of the unwanted
mass relations can be obtained by using bulk fields \cite{gutex}.

Even more interesting is the case when a higher dimensional spinor 
interacts with a non-trivial background of solitonic type. 
It has been known for a long time that this provides a mechanism to 
obtain massless four dimensional chiral fermions \cite{sol,sol2}. 
Moreover, since 
the wave functions for the zero modes of the Dirac operator are localized 
around the core of the topological defect, 
such a mechanism can play a relevant role in explaining the observed 
hierarchy in the fermion spectrum \cite{ark}.
Mass terms arise dynamically from the
overlap among fermion and Higgs wave functions.
Typically, there is an exponential mapping between the parameters
of the higher dimensional theory and the four dimensional masses
and mixing angles, so that even with parameters of order one
large hierarchies are created \cite{ark2}. 
In orbifold compactifications, solitons are simulated by
scalar fields with a non-trivial parity assignment
that forbids constant non-vanishing vacuum expectation values (VEVs).
Under certain conditions, the energy is minimized by field
configurations with a non-trivial dependence upon the compact
coordinates \cite{ggh}. 
Also in this case the zero modes of the Dirac operator in such a 
background can be chiral and localized in specific regions of
the compact space. 

In models of this sort, several zero modes can originate from 
a single higher dimensional spinor \cite{sol,sol2}. For instance, in the
model studied in ref. \cite{russi} 
there is a vortex solution that arises
in the presence of two infinite extra dimensions.     
It is possible to choose the vortex background
in such a way that the number of chiral zero modes of the 
four dimensional Dirac operator is three.
Each single six dimensional spinor gives rise to three massless
four dimensional modes with the same quantum numbers,
thus providing an elegant mechanism for understanding the fermion replica.
Recently this model has been extended to the case of compact extra
dimensions \cite{russi2}. 

In the present work we propose a model where
the different fermion generations originate from orbifold compactification, 
with a natural hierarchy among the fermion masses
and with a non-trivial mixing in flavour space. 
We consider the case of
two extra dimensions compactified on the orbifold $T^2/Z_2$,
which allows for a straightforward inclusion of localized gauge fields. 
Matter is described by vector-like six dimensional fermions
with the gauge quantum numbers of one standard model generation.
As a result, the model has neither bulk nor localized gauge anomalies.
Here we focus on a toy model with two generations, to discuss
in a simple setting the features of our proposal
and postpone the search for a fully realistic model to a future
investigation. The two generations arise as zero modes of
the Dirac operator 
by eliminating the unwanted chiralities of vector-like
six dimensional spinors through an orbifold projection.
By consistency, the fermion mass is required to be $Z_2$-odd
and, as a consequence, the two independent zero modes are
localized at the opposite sides of the sixth dimension.
The two fermion generations are distinguished by localizing
the Higgs doublet around $x_6=0$. This gives automatically 
rise to the desired mass hierarchy.
A non-trivial flavour mixing also comes out naturally and does not need 
any additional structure beyond the minimal one.
Such a mixing is related to a soft breaking of the six dimensional
parity symmetry.
In particular, in the quark sector of our toy model,
the empirical relation $\theta_C\approx \sqrt{m_d/m_s}$ can
be easily accommodated.
The essence of our proposal is to address within a unique higher dimensional 
framework both the problem of fermion replica and that of flavour 
symmetry breaking. We believe that the model described in the next sections 
represents a concrete step towards the realization of such a 
program.

%
%
\bc
{\bf 2. A model}
\ec
We want to identify the different fermion generations
with the appropriate components of a higher dimensional
fermion. A five dimensional (5D) fermion contains two 4D components with 
opposite chirality. After projecting out the wrong chirality
we are thus left with a single generation. In 6D fermions can be chiral
and a 6D chiral fermion has the same content of a 5D fermion.
The simplest conceivable case where two replica with the same
4D chirality are present is that of a 6D vector-like
fermion and we will adopt this choice to build a toy model
with two fermion generations. 
To this purpose
we consider two extra spatial dimensions compactified on
the orbifold $T^2/Z_2$, where $T^2$ is the torus defined by 
$x_i \to x_i+2 \pi R_i$ $(i=5,6)$ and $Z_2$ is the parity 
symmetry $(x_5,x_6)\to (-x_5,-x_6)$. As fundamental region of the
orbifold we can take, for instance, the rectangle 
$(\vert x_5 \vert \le \pi R_5)$, $(0\le x_6 \le \pi R_6)$ (see fig. 1).
\begin{figure}[!h]
\centerline{
\psfig{figure=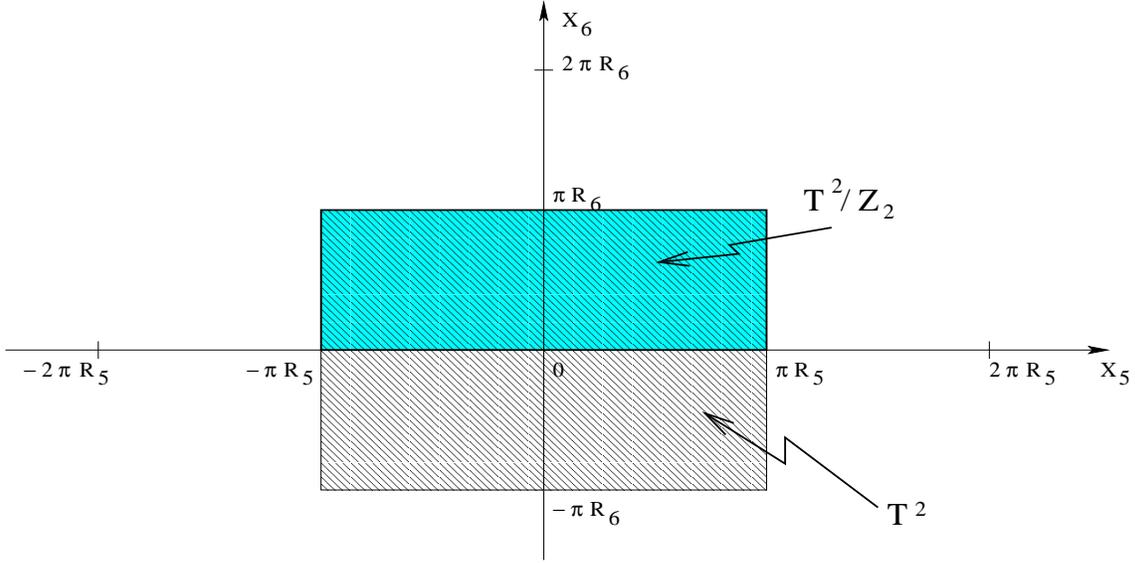,height=7.5cm,width=15cm}}
\vspace{0.3cm}
\caption{Fundamental region of the orbifold $T^2/Z_2$.
\label{orb}}
\end{figure}
There are four inequivalent fixed points under $Z_2$. In
the chosen fundamental region they can be identified
with $(x_5,x_6)=(0,0)$, $(\pi R_5,0)$,
$(0,\pi R_6)$, $(\pi R_5,\pi R_6)$.
Our theory is invariant under the gauge group 
SU(3)$\otimes$SU(2)$\otimes$U(1). 
To justify the use of 6D
vector-like fermions as building blocks of our model,
we also ask invariance under 6D parity to start with. 
As a consequence, the Lagrangian has
6D vector-like fermions $\Psi^{(\alpha)}$ $(\alpha=1,...5)$, 
one for each irreducible representation of the SM, as summarized in table 1. 
With this set of fermion fields, our model is automatically free from
6D gauge anomalies.
As we will see later on, requiring exact 6D parity symmetry is too strong
an assumption to obtain a `realistic' fermion spectrum. Although eventually
we will relax this assumption, for the time being we carry on our 
construction by enforcing 6D parity invariance.
\vspace{0.1cm}
\begin{table}[!ht]
\caption{Vector-like 6D fermions and their gauge quantum numbers.
\label{tab1}}
\vspace{0.4cm}
\begin{center}
\begin{tabular}{|c|c|c|c|}   
\hline
& & & \\                         
{\tt field} & SU(3) & SU(2) & U(1) \\ 
& & & \\
\hline
& & & \\                         
$\Psi^{(1)}$& 3& 2& +1/6\\ 
& & & \\
\hline
& & & \\                         
$\Psi^{(2)}$& 3& 1& +2/3\\ 
& & & \\
\hline
& & & \\                         
$\Psi^{(3)}$& 3& 1& -1/3\\ 
& & & \\
\hline
& & & \\                         
$\Psi^{(4)}$& 1& 2& -1/2\\ 
& & & \\
\hline
& & & \\                         
$\Psi^{(5)}$& 1& 1& -1\\ 
& & & \\
\hline
\end{tabular}
\end{center}
\end{table}
We have:
\be
{\cal L}_g={\cal L}_{gauge}+i\sum_{\alpha=1}^5 \overline{\Psi^{(\alpha)}}
\Gamma^A D_A \Psi^{(\alpha)}~~~,
\label{lg}
\ee
where ${\cal L}_{gauge}$ stands for the 6D kinetic term for
the gauge vector bosons $A_M$ $(M=0,...3,5,6)$ of 
SU(3)$\otimes$SU(2)$\otimes$U(1) and
$D_A \Psi^{(\alpha)}$ $(A=0,...3,5,6)$ denotes the appropriate 
fermion covariant derivative. We recall that, up to the
$(x_5,x_6)$ dependence, a 6D vector-like spinor is equivalent to a pair of 
4D Dirac spinors: $\Psi=(\eta,\chi)^T$. Moreover each 6D fermion can be 
split into two chiralities $\Psi=\Psi_{+}+\Psi_{-}$, eigenstates 
of $\Gamma_7$: $\Psi_{\pm}=(1\pm\Gamma_7)/2~~ \Psi$. 
We choose a representation for the Dirac matrices in 6D where
$\Gamma_7=\gamma_5\otimes\sigma_3$ (see the appendix),
where $\sigma_3$ is the third Pauli matrix, so that in terms of
4D chiralities we have: $\Psi_{+}=(\eta_R,\chi_L)^T$ and 
$\Psi_{-}=(\eta_L,\chi_R)^T$. Each component $\eta_{L,R}$, $\chi_{L,R}$
transforms in the same way under the gauge group.
All fields are assumed to be periodic in 
$x_5$ and $x_6$. By inspecting the kinetic terms, we see
that consistency with the orbifold projection requires a non-trivial 
assignment of the $Z_2$ parity. We take $A_\mu$ $(\mu=0,...3)$ 
even under $Z_2$ and $A_i$ $(i=5,6)$ $Z_2$-odd. 
In the fermion sector, $\eta_{R(L)}$ and $\chi_{R(L)}$ should have 
the same $Z_2$ parity, which should be opposite for
$\eta_{R(L)}$ and $\chi_{L(R)}$. We choose 
$Z_2(\eta^{(\alpha)}_R,\chi^{(\alpha)}_L,\eta^{(\alpha)}_L,\chi^{(\alpha)}_R)$
equal to 
$(-1,+1,+1,-1)$ for $\alpha=1,4$, and $(+1,-1,-1,+1)$ for $\alpha=2,3,5$.
At this level the zero modes are the gauge vector bosons of the
standard model and two independent chiral fermions
for each irreducible representation of the standard model,
describing two massless generations. 
There are no gauge anomalies in our model.
Bulk anomalies are absent because the 6D fermions are vector-like.
There could be gauge 4D anomalies localized at the four orbifold
fixed points \cite{orban,asa}. In our model based on
$T^2/Z_2$, the anomalies are the same at each fixed point
and they actually vanish with the quantum number
assignments of table 1
\footnote{We have explicitly checked this by adapting the analysis
described in ref. \cite{asa}.}. Indeed they are proportional
to the anomalies of the 4D zero 
modes, which form two complete fermion generations,
thus providing full 4D anomaly cancellation.
Fermion masses in six dimension and Yukawa couplings
do not modify this conclusion. 

In the absence of additional interactions, each zero mode 
is constant with respect
to $x_5$ and $x_6$. Even by introducing a 6D (parity invariant) 
Yukawa interaction
between fermions and a Higgs electroweak doublet, we do not
break the 4D flavor symmetry, which is maximal.
The first step to distinguish the two fermion generations
is to localize them in different regions of the compact space.
In our model this can be done in a very simple way, by introducing
a 6D fermion mass term
\bea
{\cal L}_m&=&\sum_{\alpha=1}^5 m_{(\alpha)} \overline{\Psi^{(\alpha)}}
\Psi^{(\alpha)}\nn\\
&=&\sum_{\alpha=1}^5 m_{(\alpha)} \overline{\Psi^{(\alpha)}}
\left(\frac{1-\Gamma_7}{2}\right)\Psi^{(\alpha)}+h.c.~~~,
\label{lm}
\eea
where 6D parity requires $m_{(\alpha)}$ to be real.
This term is gauge invariant and relates left and right 4D
chiralities. Therefore the mass parameters $m_{(\alpha)}$ are
required to be $Z_2$-odd and cannot be constant in the whole 
$(x_5,x_6)$ plane. 
The simplest possible choice for 
$m_{(\alpha)}$ is a constant in the orbifold fundamental region
\footnote{Of course there is not a unique way of choosing the
fundamental region and this leads to several possible choices
for $m_{(\alpha)}$. Although we are now regarding $\mu_{(\alpha)}$
as real parameters, in the next section we will also need 
results for complex $\mu_{(\alpha)}$. For this reason we carry out 
our analysis directly in the complex case.}:
\be
m_{(\alpha)}(x_5,x_6)=\mu_{(\alpha)} \epsilon(x_6)~~~,
\ee 
where $\epsilon(x_6)$ denotes the (periodic) sign function.
This function can be regarded as a background field.
In a more fundamental theory it could arise dynamically
from the VEV of a gauge singlet scalar field, periodic and $Z_2$-odd 
\cite{ggh}.
Then the parameters $\mu_{(\alpha)}$ would essentially represent
Yukawa couplings. 
In our toy model we regard $\epsilon(x_6)$ as an external
fixed background and neglect its dynamics.

The properties of the 4D light fermions are now described by
the zero modes of the 4D Dirac operator in the background 
proportional to $\epsilon(x_6)$. These zero modes are
the normalized solutions to the differential equations:
\bea
(\partial_5+i\partial_6) \chi^{(\alpha)}_L+ \mu_{(\alpha)} \epsilon(x_6) 
\eta^{(\alpha)}_L &=&0\nn\\
(\partial_5-i\partial_6) \eta^{(\alpha)}_L+ \mu_{(\alpha)}^*\epsilon(x_6) 
\chi^{(\alpha)}_L &=&0\nn\\
-(\partial_5+i\partial_6) \chi^{(\alpha)}_R+\mu_{(\alpha)}^*\epsilon(x_6) 
\eta^{(\alpha)}_R &=&0\nn\\
-(\partial_5-i\partial_6) \eta^{(\alpha)}_R+ \mu_{(\alpha)}\epsilon(x_6) 
\chi^{(\alpha)}_R &=&0~~~,
\label{fstorder}
\eea
with periodic boundary conditions for all fields and with
the $Z_2$ parities defined above. By applying standard techniques 
(see appendix) we obtain:
\begin{itemize}
\item{$\alpha=1,4$}
\end{itemize}
\bea
\left(
\begin{array}{c}
\eta^{(\alpha)}_R\\
\chi^{(\alpha)}_R
\end{array}
\right)&=&0\nn\\
\left(
\begin{array}{c}
\eta^{(\alpha)}_L\\
\chi^{(\alpha)}_L
\end{array}
\right)&=&
f^{(\alpha)}_1(x)
\left(
\begin{array}{c}
1\\
i \dd\frac{\mu_{(\alpha)}}{\vert\mu_{(\alpha)}\vert}
\end{array}
\right)
\xi^{(\alpha)}_1(x_5,x_6)
+
f^{(\alpha)}_2(x)
\left(
\begin{array}{c}
1\\
- i \dd\frac{\mu_{(\alpha)}}{\vert\mu_{(\alpha)}\vert}
\end{array}
\right)
\xi^{(\alpha)}_2(x_5,x_6)~~~,
\label{zm14}
\eea
\begin{itemize}
\item{$\alpha=2,3,5$}
\end{itemize}
\bea
\left(
\begin{array}{c}
\eta^{(\alpha)}_R\\
\chi^{(\alpha)}_R
\end{array}
\right)&=&
f^{(\alpha)}_1(x)
\left(
\begin{array}{c}
1\\
-i \dd\frac{\mu_{(\alpha)}^{~*}}{\vert\mu_{(\alpha)}\vert}
\end{array}
\right)
\xi^{(\alpha)}_1(x_5,x_6)
+
f^{(\alpha)}_2(x)
\left(
\begin{array}{c}
1\\
i \dd\frac{\mu_{(\alpha)}^{~*}}{\vert\mu_{(\alpha)}\vert}
\end{array}
\right)
\xi^{(\alpha)}_2(x_5,x_6)\nn\\
\left(
\begin{array}{c}
\eta^{(\alpha)}_L\\
\chi^{(\alpha)}_L
\end{array}
\right)&=&0~~~,
\label{zm235}
\eea
where $f^{(\alpha)}_{1,2}(x)$ are 4D chiral spinors:
\be
\begin{array}{cc}
f^{(1)}_1=
\left(
\begin{array}{c}
u_L\\
d_L
\end{array}
\right)&
f^{(1)}_2=
\left(
\begin{array}{c}
c_L\\
s_L
\end{array}
\right)\nn\\
f^{(2)}_1=u_R& f^{(2)}_2=c_R\nn\\
f^{(3)}_1=d_R& f^{(3)}_2=s_R\nn\\
f^{(4)}_1=
\left(
\begin{array}{c}
\nu_{e L}\\
e_L
\end{array}
\right)&
f^{(4)}_2=
\left(
\begin{array}{c}
\nu_{\mu L}\\
\mu_L
\end{array}
\right)\nn\\
f^{(5)}_1=e_R& f^{(5)}_2=\mu_R
\end{array}~~~,
\ee
whereas $\xi^{(\alpha)}_{1,2}(x_5,x_6)$ are functions describing
the localization of the zero modes in the compact space:
\bea
\xi^{(\alpha)}_1(x_5,x_6)&=&
\frac{e^{\dd{-\pi \vert\mu_{(\alpha)}\vert R_6}}}{\sqrt{2\pi R_5}}~
\frac{\sqrt{\vert\mu_{(\alpha)}\vert}}{\sqrt{1-e^{\dd{-2\pi\vert\mu_{(\alpha)}\vert R_6}}}}~~
e^{~\dd{\vert \mu_{(\alpha)} x_6\vert_{per}}}\nn\\
\xi^{(\alpha)}_2(x_5,x_6)&=&
\frac{1}{\sqrt{2\pi R_5}}~
\frac{\sqrt{\vert\mu_{(\alpha)}\vert}}{\sqrt{1-e^{\dd{-2\pi\vert\mu_{(\alpha)}\vert R_6}}}}~~
e^{\dd{-\vert \mu_{(\alpha)} 
x_6\vert_{per}}}~~~.
\label{xi}
\eea
In the above equations $\vert x_6\vert_{per}$ denotes a periodic
function, coinciding with the ordinary $\vert x_6\vert$
in the interval $[-\pi R_6,\pi R_6]$. As in the case $m_{(\alpha)}=0$, 
for each 6D spinor we have two independent chiral zero modes, 
whose 4D dependence is described by $f^{(\alpha)}_{1,2}$. 
They are still constant in $x_5$, but not
in $x_6$. 
\begin{figure}[!t]
\centerline{
\psfig{figure=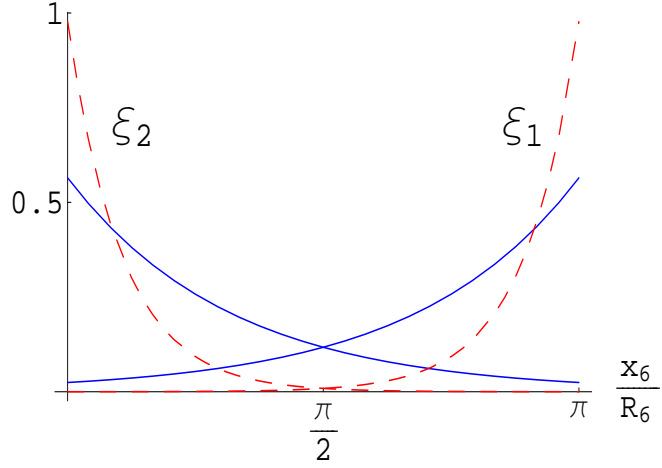,height=7.5cm,width=10cm}}
\vspace{0.3cm}
\caption{Wave functions $\xi^{(\alpha)}_{1,2}(x_5,x_6)$, 
in units of $\sqrt{R_5 R_6}$. 
They have been obtained by choosing $\vert\mu_{(\alpha)}\vert R_6=1(3)$
for continuous-blue (dashed-red) lines.
\label{zeromodes}}
\end{figure}
Indeed, the zero mode proportional to $f^{(\alpha)}_2$ 
is localized at $x_6=0$ (mod $2\pi R_6$), whereas that proportional to 
$f^{(\alpha)}_1$ is peaked around $x_6=\pi R_6$ (mod $2\pi R_6$)
(see figure 2).
The two zero modes with well-defined localization properties
in the compact space have non-trivial components both
along $\eta$ and along $\chi$ and they are orthogonal
to each other. The constant factors in eqs. (\ref{xi})
normalize the zero modes to 1. In our toy model, the number of zero modes
is not related to a non-trivial topological property
of the background $\epsilon(x_6)$. The two zero modes are
determined by the orbifold projection. The presence of the
background only induces a separation of the corresponding
wave functions in the compact space. Actually we can go smoothly
from localized to constant wave functions, by turning off
the constants $\mu_{(\alpha)}$, as apparent from eqs. (\ref{xi}).

With the introduction of the background, we now have 
two fermion generations, one sat at $x_6=0$ and the other
at $x_6=\pi R_6$. From the point of view of the four-dimensional observer,
who cannot resolve distances in the extra space, 
there is still a maximal flavour symmetry and, indeed,
all fermions are still massless at this level.
Fermions can acquire masses in the usual way,
by breaking the electroweak symmetry via the non-vanishing VEV
of a Higgs doublet $H$.
If such a VEV were a constant in $x_6$, then we would
obtain equal masses for the two fermion generations.
Thus, to break the 4D flavour symmetry we need a non-trivial
dependence of the Higgs VEV upon $x_6$. There are several
ways to achieve this. For instance, we might assume that
$H$ is a bulk field. Under certain conditions it may happen that
the minimum of the energy is no longer $x_6$-constant. 
Examples of this kind are well-known
in the literature \cite{ncvev}. 
If $H$ interacts with a suitable $x_6$-dependent 
background, there is a competition between the
kinetic energy term, which prefers constant configurations,
and the potential energy term, which may favour a $x_6$-varying
VEV. In non-vanishing portions of the parameter space
the minimum of the energy can depend non-trivially on $x_6$.  
In the minimal version of our toy model we will  
simulate this dependence in the simplest
possible way, by introducing a Higgs doublet $H$ with hypercharge +1/2 
localized along the line $x_6=0$
\footnote{Alternatively, we could assume that $H$ is localized
at the orbifold fix point $(x_5,x_6)=(0,0)$. From the point of
view of fermion masses and mixing angles, the two choices
are equivalent. To avoid singular terms in the action, we 
could also consider a mild localization, described by some
smooth limit of the Dirac delta functions involved in the
present treatment. Our results would not be qualitatively affected.}.
The most
general Yukawa interaction term invariant under $Z_2$, 6D parity
and SU(3) $\otimes$ SU(2) $\otimes$ U(1) reads:
\be
{\cal L}_Y=
\left[
y_u~ \tilde{H}^\dagger~
\overline{\Psi^{(2)}}\Psi^{(1)}
+
y_d~ H^\dagger~
\overline{\Psi^{(3)}}\Psi^{(1)}
+
y_e~ H^\dagger~
\overline{\Psi^{(5)}}\Psi^{(4)}
+h.c.\right]
\delta(x_6)~~~,
\label{lY}
\ee
where $\tilde{H}=i\sigma^2 H^*$. Notice that $H$ has dimension
+3/2 and $y$ has dimension -3/2, in mass units.
In the next section we will
see how a realistic pattern of masses and mixing angles
arises from these Yukawa interactions.

Summarizing, our model is described by the Lagrangian:
\be
{\cal L}={\cal L}_g+{\cal L}_m+{\cal L}_Y+{\cal L}_H~~~,
\ee
where ${\cal L}_g$, ${\cal L}_m$, ${\cal L}_Y$ are given in
eqs. (\ref{lg}), (\ref{lm}) and (\ref{lY}), respectively,
while ${\cal L}_H$, localized at $x_6=0$, 
contains the kinetic term for the Higgs
doublet and the scalar potential that breaks spontaneously
SU(2)$\otimes$U(1).
The complex phases in $y$ can be completely eliminated via 
field redefinitions: in the limit of exact 6D parity symmetry
all parameters are real.
%
%
\bc
{\bf 3. Masses and Mixing Angles}
\ec
The fermion mass terms arise from ${\cal L}_Y$ after electroweak
symmetry breaking, here described by 
$\langle H\rangle=(0~ v/\sqrt{2})^T$. 
To evaluate the fermion mass
matrices we should expand the 6D fermion fields in 4D modes
and then perform the $x_5$ and $x_6$ integrations. In practice,
if we focus on the lightest sector, we can keep 
only the zero modes in the expansion. We obtain:
\bea
m_u&=&
\frac{y_u}{\sqrt{2}}~ v~ 
\frac{\sqrt{\vert \mu_{(1)} \mu_{(2)} \vert}}{2\sqrt{(1-\lambda_1^2)
(1-\lambda_2^2)}}
\left(
\begin{array}{cc}
c_{u-}~ \lambda_1 \lambda_2 & c_{u+}~ \lambda_2\\
c_{u+}~ \lambda_1 & c_{u-}
\end{array}
\right)\nn\\
m_d&=&
\frac{y_d}{\sqrt{2}}~ v~ 
\frac{\sqrt{\vert \mu_{(1)} \mu_{(3)} \vert}}{2\sqrt{(1-\lambda_1^2)
(1-\lambda_3^2)}}
\left(
\begin{array}{cc}
c_{d-}~ \lambda_1 \lambda_3 & c_{d+}~ \lambda_3\\
c_{d+}~ \lambda_1 & c_{d-}
\end{array}
\right)\nn\\
m_e&=&
\frac{y_e}{\sqrt{2}}~ v~ 
\frac{\sqrt{\vert \mu_{(4)} \mu_{(5)} \vert}}{2\sqrt{(1-\lambda_4^2)
(1-\lambda_5^2)}}
\left(
\begin{array}{cc}
c_{e-}~ \lambda_4 \lambda_5 & c_{e+}~ \lambda_5\\
c_{e+}~ \lambda_4 &c_{e-} 
\end{array}
\right)~~~,
\eea
where 
\be
c_{u\pm}=1\pm\frac{\mu_{(1)}\mu_{(2)}}{\vert
\mu_{(1)}\mu_{(2)}\vert}~~~,~~~~~
c_{d\pm}=1\pm\frac{\mu_{(1)}\mu_{(3)}}{\vert
\mu_{(1)}\mu_{(3)}\vert}~~~,~~~~~
c_{e\pm}=1\pm\frac{\mu_{(4)}\mu_{(5)}}{\vert
\mu_{(4)}\mu_{(5)}\vert}~~~,
\ee
and
\be
\lambda_\alpha=e^{\dd{-\pi\vert\mu_{(\alpha)}\vert R_6}}~~~.
\ee
These mass matrices, here given in the convention $\overline{f_R} m_f
f_L$, are not hermitian. It is interesting to see that,
for generic order-one values of the dimensionless combinations
$c_{f\pm}$ and $\mu_{(\alpha)} R_6$, the mass matrices 
display a clear hierarchical pattern. 
Fermion masses of the first generation are suppressed by
$\lambda_{(\alpha)}\lambda_{(\beta)}$ compared to those of the second
generation and mixing angles are of order $\lambda_{(\alpha)}$ or 
$\lambda_{(\beta)}$. This is
quite similar to what obtained in 4D models with a spontaneously
broken flavour symmetry. 
Here the role of small expansion
parameters is played by the quantities $\lambda_\alpha$.
However in our parity invariant model, the parameters $\mu_{(\alpha)}$
are real and the coefficients $c_{f\pm}$ are `quantized'. Either
$c_{f+}$ or $c_{f-}$ should vanish and this implies no mixing.
Indeed when 6D parity is conserved, we have only two possible
orientations of the fermionic zero modes in the $(\eta,\chi)$
space: either $(1,i)$ or $(1,-i)$, as apparent from eqs. (\ref{zm14})
and (\ref{zm235}). Thus the scalar product between two zero modes
in the $(\eta,\chi)$ space is either maximal or zero.
Modulo a relabelling among first and second generations, this gives
rise to a perfect alignment of mass matrices and a vanishing overall mixing.
To overcome this problem, we should relax the assumption of
exact 6D parity symmetry 
\footnote{There are other possibilities that lead to a non-vanishing
mixing. For instance we could introduce several independent backgrounds
and couple them selectively to the different fermion fields.
In our view, the solution discussed in the text is the simplest one.}. 
We will assume that 6D parity is broken
`softly', by the fermion-background interaction described by ${\cal L}_m$.
This can be achieved by taking complex values for the
mass coefficients $\mu_{(\alpha)}$
\footnote{All previous equations remain unchanged, but the first equality
in eq. (\ref{lm}). Only the second one is correct.}. 
In a fundamental theory such a breaking could be spontaneous: if $m_{(\alpha)}$
were complex fields, then the Lagrangian would still be invariant under 
6D parity acting as 
$m_{(\alpha)} \leftrightarrow m_{(\alpha)}^\dagger$.
It might occur that the dynamics of the fields $m_{(\alpha)}$
led to complex VEVs for $m_{(\alpha)}$,
thus spontaneously breaking parity. In our toy model we 
will simply assume the existence of such a complex background.
All the relations that we have derived hold true for the complex case 
as well and we have now hierarchical mass matrices with a non-trivial 
intergenerational mixing.
By expanding the results at leading order in $\lambda_\alpha$ we find:
\be
m_c=\vert y_u \vert v \frac{\sqrt{\vert \mu_{(1)}\mu_{(2)}
\vert}}{2\sqrt{2}} \vert c_{u-} \vert~~~~~~
m_s=\vert y_d \vert v \frac{\sqrt{\vert \mu_{(1)}\mu_{(3)}
\vert}}{2\sqrt{2}} \vert c_{d-} \vert~~~~~~
m_\mu=\vert y_e \vert v \frac{\sqrt{\vert \mu_{(4)}\mu_{(5)}
\vert}}{2\sqrt{2}} \vert c_{e-} \vert~~,
\ee
and
\be
\frac{m_u}{m_c}=\frac{\vert c_{u+}^2 - c_{u-}^2 \vert}{\vert c_{u-}
\vert^2} \lambda_1 \lambda_2~~~~~
\frac{m_d}{m_s}=\frac{\vert c_{d+}^2 - c_{d-}^2 \vert}{\vert c_{d-}
\vert^2} \lambda_1 \lambda_3~~~~~
\frac{m_e}{m_\mu}=\frac{\vert c_{e+}^2 - c_{e-}^2 \vert}{\vert c_{e-}
\vert^2} \lambda_4 \lambda_5~~~.
\label{mom}
\ee
Finally, after absorbing residual phases in the definition
of the $s$ and $c$ 4D fields, the matrices $m_u^\dagger m_u$ and 
$m_d^\dagger m_d$ are diagonalized by orthogonal transformations
characterized by mixing angles $\theta_{u,d}$:
\be
\theta_{u,d}= \left\vert\frac{c_{u,d+}}{c_{u,d-}}\right\vert
\lambda_1~~~,
\label{ma}
\ee
still at leading order in $\lambda_\alpha$.
Therefore the Cabibbo angle is given by:
\be
\theta_C=\left(\left\vert\frac{c_{d+}}{c_{d-}}\right\vert-
\left\vert\frac{c_{u+}}{c_{u-}}\right\vert\right)\lambda_1~~~.
\label{ca}
\ee
Barring accidental cancellations in the relevant combinations
of the coefficients $c_{f\pm}$, the Cabibbo angle is of order $\lambda_1$.
Then, by assuming $\lambda_3\approx\lambda_1$ and 
$\lambda_2\approx\lambda_1^3$ we reproduce the correct order 
of magnitude of mass ratios in the quark sector.
These are small numbers in the 4D theory, but can be obtained
quite naturally from the 6D point of view: $\mu_{(1)} R_6
\approx \mu_{(3)} R_6\approx 0.5$ and $\mu_{(2)} R_6\approx 1.3$.
Similarly, by taking $\lambda_4\lambda_5\approx\lambda_1^2$
we can naturally fit the lepton mass ratio.

It can be useful to comment about the way flavour symmetry
is broken in this toy model. Before the introduction of the 
Yukawa interactions and modulo U(1) anomalies, the flavour symmetry group
is U(2)$^5$. After turning the Yukawa couplings on, we can 
consider several limits. When $R_6\to \infty$, the quantities
$\lambda_{(\alpha)}$ vanish and the flavour symmetry is broken
down to U(1)$^5$, acting non-trivially on the lightest sector.
If $R_6$ is finite and non-vanishing, U(1)$^5$ is in turn
completely broken down by $\lambda_{(\alpha)}\ne 0$.
Nevertheless, contrary to what happens in models with abelian
flavour symmetries, the coefficients of order one that
multiply the symmetry breaking parameters $\lambda_{(\alpha)}$
are now related one to each other. This can be appreciated by
taking the limit $R_6\to 0$. We have $\lambda_{(\alpha)}=1$ and the residual 
flavour symmetry is a permutation symmetry, separately for the lepton
and the quark sectors: $S_2\otimes S_2$.

Let us now briefly comment about neutrino masses and mixings in this
set-up.  The most straightforward way to produce neutrino masses is to
add a gauge singlet 6D fermion field, $\Psi^{(6)}$, with $Z_2$
assignments $(+1,-1,-1,+1)$. As for the case of charged fermions, by
introducing a mass term for $\Psi^{(6)}$ as in eq. (\ref{lm}) and a
Yukawa interaction with $\Psi^{(4)}$ and $\tilde H$ as in eq.
(\ref{lY}), we obtain a Dirac neutrino mass term \be m_\nu =
\frac{y_\nu}{\sqrt{2}} v \frac{\sqrt{\vert\mu_{(4)} \mu_{(6)}\vert
}}{2\sqrt{ (1- \lambda_4^2 ) (1- \lambda_6^2)}} \left( \matrix{ c_{\nu
-} \lambda_4 \lambda_6 & c_{\nu +} \lambda_6 \cr c_{\nu +} \lambda_4 &
c_{\nu -}} \right) \ee A large mixing angle in the leptonic sector,
$\theta_L$, is obtained for $\lambda_4 = O(1)$, in which case the
neutrino mass hierarchy, \be \frac{m_{\nu_1}}{m_{\nu_2}} =
\frac{|c_{\nu +}^2 - c_{\nu -}^2 |}{|c_{\nu -}|^2} \lambda_4 \lambda_6
~~~, \ee is controlled by $\lambda_6$.  At leading order, the left
mixings in $m_e^\dagger m_e$ and $m_\nu^\dagger m_\nu$ correspond to
\be \tan 2\theta_{e,\nu} = \frac{2~ c_{e,\nu +}~c_{e,\nu
-}~\lambda_4}{(c_{e,\nu -}^2-c_{e,\nu +}^2~\lambda_4^2)}~~~, \ee so
that $\theta_L \equiv \theta_\nu-\theta_e$ is naturally large.  As in
4D, the smallness of these Dirac neutrino masses with respect to the
electroweak scale has to be imposed by an ad hoc suppression of the
Yukawa couplings $y_\nu$.  A natural suppression could be achieved by
considering also a Majorana mass term in the bulk for the field
$\Psi^{(6)}$.  Alternatively, one could write localized Majorana mass
terms directly on the 4D brane or even exploit a seventh warped
extra-dimension \cite{app}.
%
%
\bc
{\bf 4. Which scale for flavour physics?}
\ec
Our 6D toy model is non renormalizable. It is characterized
by some typical mass scale $\Lambda$. At energies larger than
this typical scale, the description offered by the model
is not accurate enough and some other theory should replace it.
Up to now we have not specified $\Lambda$.
We could have in mind a traditional picture where $\Lambda$
is very large, perhaps close to the 4D Planck scale,
where presumably all particle interactions, including
the gravitational one, are unified in a fundamental theory.
In this scenario we have the usual hierarchy problem.
Clearly our simple model cannot explain why $v<<\Lambda^{3/2}$ 
and we should rely on some additional mechanism to render
the electroweak breaking scale much smaller compared to $\Lambda$. 
A supersymmetric or warped version of our toy model could 
alleviate the technical aspect of the hierarchy problem.
Alternatively, we could ask how small could $\Lambda$ be
without producing a conflict with experimental data.
For simplicity we assume that the two radii $R_5$, $R_6$ are
approximately of the same order $R$. Due to the different dimension
between 6D and 4D fields, coupling constants of the effective 4D
theory are suppressed by volume factors and we require 
$\Lambda R\ge1$ to work in a weakly coupled regime.
Therefore, lower bounds on $1/R$ are also lower bounds for
$\Lambda$. Lower bounds on $1/R$ come 
from the search of the first Kaluza-Klein
modes at the existing colliders or from indirect effects
induced by the additional heavy modes \cite{orbmod,orb6d,preh,dela}.
These last effects lead to departures from the SM predictions
in electroweak observables. From the precision tests
of the electroweak sector, we get a lower bound on $1/R$ in the $TeV$ range.
However, the most dangerous indirect effects are those 
leading to violations of universality in gauge interactions
and those contributing to flavour changing processes.
Indeed, whenever we have a source of flavour symmetry breaking,
we expect a violation of universality at some level.
In the SM such violation comes through loop effects from the Yukawa couplings
and it is tiny. In our model, as we will see,
such effects can already arise at tree level and, to respect the 
experimental bounds, a sufficiently large scale $1/R$ is needed \cite{bur}. 

Since in each fermion sector the two generations are described by two copies
of the same wave function, differing only in their localization
along $x_6$, the universality of the gauge interactions
will be guaranteed if the gauge vector bosons have a wave
function perfectly constant in $x_6$.
This is the case only for massless gauge vector bosons, such as the photon,
but, as we will see now, not necessarily for the massive gauge vector 
bosons like $W$ and $Z$. 
Moreover, also the higher Kaluza-Klein modes of all 
gauge bosons have non-constant wave functions and their interactions
with split fermions are in general non-universal. 

We start by discussing the interactions between the lightest fermion
generations and the observed $W$ and $Z$ vector bosons.
Consider, for simplicity, the limit of vanishing
gauge coupling $g'$ for $U(1)$. Then the free equation of
motion for the gauge bosons $W_\mu$ of $SU(2)$ reads:
\be
\Box W_\mu + \frac{g^2}{2} h^2(x_6) W_\mu=0~~~,
\label{gvb}
\ee
where $h(x_6)$ denotes the $x_6$-dependent VEV of the 
Higgs doublet $H$. To avoid problems in dealing with
singular, ill-defined functions, here $h(x_6)$ is
a smooth function, VEV of a 6D bulk field. From the eq. 
(\ref{gvb}) we will see that,
if $h(x_6)$ is not constant, then the lightest mode
for the gauge vector bosons is no longer described
by a constant wave function. Therefore the 4D gauge interactions,
resulting from the overlap of fermion and vector bosons
wave functions, can be different for the two generations.

In general we are not able to solve the above equation exactly,
but we can do this by a perturbative expansion in $g^2$,
which we could justify a posteriori.
At zeroth order the $W^3$ mass and the corresponding
wave function are given by:
\be
(m^{(0)}_W)^2=0~~~~~~~~~~~~~
W^{(0)}_\mu=\frac{1}{\sqrt{2 \pi^2 R_5 R_6}}~~~.
\ee
At first order we find:
\bea
m_W^2&=&\frac{g^2}{2\pi R_6} \int_{0}^{+\pi R_6} dx_6~ h^2(x_6)\nn\\
W_\mu&=&W^{(0)}_\mu(1+\delta W_\mu(x_6))\nn\\
\delta W_\mu(x_6)&=&\int_0^{x_6}du\int_0^u dz(\frac{g^2}{2} h^2(z)-m_W^2)~~~,
\label{wf}
\eea
modulo an arbitrary additive constant in $W_\mu$, that can be 
adjusted by normalization. We see that when $h(x_6)$ is constant,
the usual result is reproduced: $m_W^2=g^2 h^2/2$ and the 
corresponding wave function does not depend on $x_6$.
Eq. (\ref{wf}) allows us to compute the fractional difference
$(g_1-g_2)/(g_1+g_2)$ between the SU(2) couplings to the first and
second fermion generation, respectively. 
Focusing on $W^3_\mu$, we obtain:
\be
\left\vert\frac{g_1-g_2}{g_1+g_2}\right\vert=
\frac{\int_0^{\pi R_6}dx_6~
\left(\vert\xi^{(\alpha)}_1\vert^2-\vert\xi^{(\alpha)}_2\vert^2\right)
\delta W^3_\mu}
{\int_0^{\pi R_6}dx_6~
\left(\vert\xi^{(\alpha)}_1\vert^2+\vert\xi^{(\alpha)}_2\vert^2\right)}~~~,
\label{rg}
\ee where $\alpha=1,4$.  As expected, if $\delta W^3_\mu$ is
$x_6$-constant, then the gauge couplings are universal. From the
precision tests of the SM performed in the last decade at LEP and SLC
we expect that such a difference should not exceed, say, the per-mill
level.  We have analyzed numerically eq. (\ref{rg}) for several
choices of the parameters and for several possible profiles of the VEV
$h(x_6)$. We found that universality is respected at the per-mill
level for $m_W^2 R_6^2<O(10^{-3})$ or $1/R_6>3~TeV$.

Much more severe are the bounds associated to the
interactions of the higher modes
arising from the Kaluza-Klein decomposition of the gauge vector bosons:
\be
A_\mu=i \sum_{c,m,n} t^c A_\mu^{c (m,n)}(x) z_{mn}(x_5,x_6)~~~,
\label{kkg}
\ee
where $t^c$ are the generators of the gauge group factor,
$A_\mu^{c (m,n)}(x)$ the corresponding 4D vector bosons and
$z_{mn}(x_5,x_6)$ the periodic, $Z_2$-even wave functions:
\be
z_{mn}(x_5,x_6)=\frac{1}{\sqrt{\pi^2 R_5 R_6 2^{\delta_{m,0}\delta_{n,0}}}}
\cos(m\frac{x_5}{R_5}+n\frac{x_6}{R_6})~~~.
\ee
In eq. (\ref{kkg}) $m$ and $n\ge 0$ are integers: 
$m$ runs from $-\infty$ to $+\infty$ for positive $n$ and from $0$ 
to $+\infty$ for $n=0$. From eq. (\ref{lg})
we obtain the 4D interaction term:
\be
-\frac{g}{\Lambda} \sum_{a,b}\sum_{m,n} c_{ab}^{mn}~ 
\overline{f^{(\alpha)}_a(x)}~\gamma^\mu t^c~ f^{(\alpha)}_b(x)~ 
A_\mu^{c (m,n)}(x)~~~,
\label{hint}
\ee
where $g$ denotes the gauge coupling constant of the relevant group
factor and the scale $\Lambda$ has been included to make $g$
dimensionless. The coefficients $c_{ab}^{mn}$ $(a,b=1,2)$ describe the overlap
among the fermion and gauge-boson wave functions. We obtain:
\bea
c_{ab}^{mn}&=&0~~~~~~~~~~m\ne 0\nn\\
c_{ab}^{0n}&=&0~~~~~~~~~~a\ne b\nn\\
c_{11}^{0n}&=&
\frac{1}{\sqrt{\pi^2 R_5 R_6 2^{\delta_{n,0}}}}
            \frac{4 \vert\mu_{\alpha}\vert^2 R_6^2}
{n^2 + 4 \vert\mu_{\alpha}\vert^2 R_6^2} 
            \frac{(1-(-1)^n e^{2|\mu_{\alpha}| \pi R_6})}
                 {(1- e^{2|\mu_{\alpha}| \pi R_6})}\nn\\
c_{22}^{0n}&=&(-1)^n c_{11}^{0n}
\label{hcoef}
\eea
For odd $n$, the interactions mediated by $A_\mu^{c (0,n)}$ 
are non-universal. By asking that universality holds within the 
experimental limits, we get a lower bound on $1/R$ similar
to that discussed before, of the order of some $TeV$.
However, stronger bounds are obtained from the interactions in eqs.
(\ref{hint},\ref{hcoef}), by considering their contribution
to flavour changing processes. Indeed, after electroweak symmetry
breaking, we should account for the unitary transformations bringing fermions
from the interaction basis to the mass eigenstate basis.
The terms involving $A_\mu^{c (0,2n+1)}$
are not invariant under such transformations and flavour changing
interactions are produced. By integrating out the heavy modes
$A_\mu^{c (0,2n+1)}$ we obtain an effective, low-energy
description of flavour violation in terms of four-fermion operators,
suppressed by the square of the compactification scale, $(1/R_6)^2$. 
The most relevant effects of these operators have been discussed 
by Delgado, Pomarol and Quiros 
in ref. \cite{dpq} in a context which is very close to the one 
we are considering here. By analyzing the contribution to $\Delta m_K$
and to $\epsilon_K$, these authors derived a lower bound on $1/R$
of $O(100~TeV)$ and of $O(1000~TeV)$, respectively, which
at least as an order of magnitude applies also to our model.
%
%
%
\bc {\bf 5. Outlook} \ec We have presented and analyzed a model for
flavour in two extra dimensions.  For each irreducible representation
of the SM, there is a single, vector-like 6D fermion.  Four
dimensional fermion generations arise from orbifold projection in the
compactification mechanism.  In the toy model discussed here there is
only room for two generations and a natural question is whether this
approach can be generalized to the realistic case of three
generations.  The obvious objection is that the number of 4D
components of a higher dimensional fermion is a power of two. In
principle, an odd number of massless modes can be obtained by
eliminating some of the unwanted components via orbifold projection
and/or non-periodic boundary conditions.  In our model flavour
symmetry is broken in two steps. First, the independent zero modes are
localized at different points along the sixth dimension by means of a
generalized 6D mass term, described by a scalar background.  Our
background is topologically trivial and does not modify the number of
zero modes, which is fixed by the orbifold projection.  However, the
presence of a background with a non-trivial topology may change the
number of chiral zero modes, thus contributing to reproduce the
realistic case \cite{russi,russi2}.  The flavour symmetry is then
broken by turning on standard Yukawa interaction with a Higgs field
developing a non-constant VEV.  In the model explored here the
geometry of the compact space is the simplest one and it is
essentially one dimensional. The generations are localized along the
sixth dimension, and the fifth dimension does not play any active
role.  In a more realistic model it might be necessary to fully
exploit the geometry of the compact space, in order to obtain a
successful arrangement for the zero modes.  A specific problem is
represented by the neutrino sector, that we have only briefly touched
here.  Another unpleasant feature of our toy model is that, despite
its simplicity, it contains too many parameters and there are no
testable predictions.  Clearly the issue of predictability is crucial
for a realistic model.  It is possible that, by going to the realistic
case of three generations, the number of parameters does not increase,
thus allowing for quantitative tests of this approach. Alternatively,
we could consider more constrained frameworks. A possibility could be
to exploit a grand unification symmetry to limit the number of
parameters in the fermion sector. Another interesting case is
represented by theories where the Higgs fields are identified with the
extra components of higher dimensional gauge vector bosons
\cite{hgvb}.  One of the main problems of these models is precisely to
break flavour symmetry, starting from universal Yukawa couplings,
universality being dictated by gauge invariance \cite{hgvbf}.  Our
approach could provide a possible mechanism to realize such breaking.

Despite the fact that our model is incomplete in many respects, we 
think that it possesses several interesting theoretical properties.
At variance with most of the existing 4D models, the problem of 
flavour symmetry breaking is here tightly related to the
problem of obtaining the right number of generations.
Starting from dimensionless parameters of order one,
we were able to obtain a hierarchical pattern of masses.
They are described by mass matrices that are very close in structure
to those obtained in 4D models by enforcing abelian flavour symmetries,
and provide a successful description of both quark and lepton
spectra.
The crucial difference is that, whereas in the 4D case, the order-one
coefficients multiplying powers of the symmetry breaking parameters
are completely undetermined, in our case those coefficients are
strongly correlated and predictable in terms of the underlying 
parameters.
We have also a quite non-standard interpretation of the
intergenerational mixing, that appears to be related 
to a soft breaking of 6D parity symmetry.
Starting from vector-like 6D fermions transforming as
a SM generation, we automatically obtain
cancellation of bulk and localized gauge anomalies,
a rather non-trivial result in 6D gauge theories.
Hopefully some of these features could also become part
of a more realistic framework.

\vspace*{1.0cm}
{\bf Acknowledgements.}
We thank Guido Altarelli, Riccardo Barbieri, Augusto Sagnotti, Jose Santiago
and Angel Uranga for valuable discussions. 
C.B., F.F. and I.M. thank the CERN theoretical division
for hospitality in summer/fall 2002, when we began the present work.
This project is partially
supported by the European Programs HPRN-CT-2000-00148 and HPRN-CT-2000-00149.
\vfill

\newpage

\section*{Appendix}
\vspace{0.3cm}


\subsection*{$\Gamma$ matrices}

We work with the metric
\be
\eta_{MN} = \textrm{diag}(1, -1, -1, -1, -1, -1)
\ee
where $M,N = 0,1,2,3,5,6$. 
The representation of 6D $\Gamma$-matrices we use in the text is
\be
\Gamma^{\mu} = \left(\begin{array}{cc}
\gamma^{\mu} & 0\\
0 & \gamma^{\mu}
\end{array}\right)\ ,\ \
\Gamma^5 = i\ \left(\begin{array}{cc}
0 & \gamma_5\\
\gamma_5 & 0
\end{array}\right)\ ,\ \
\Gamma^6 = i\ \left(\begin{array}{cc}
0 & i\ \gamma_5\\
-i\ \gamma_5 & 0
\end{array}\right)\ ,
\ee
with $\mu = 0,1,2,3$. Here $\gamma^{\mu}$, $\gamma_5$ are 4D $\gamma$-matrices given by
\be
\gamma^0 = \left(\begin{array}{cc}
\mathbf{1} & 0 \\
0 & \mathbf{-1}
\end{array}\right)\ ,\ \
\gamma^i = \left(\begin{array}{cc}
0 & \sigma^i \\
-\sigma^i & 0 
\end{array}\right)\ ,\ \
\gamma^5 = \left(\begin{array}{cc}
0 & \mathbf{1} \\
\mathbf{1} & 0 
\end{array}\right)\ ,
\ee
where $\sigma^i$ are the Pauli matrices.

In 6D the analogous of $\gamma_5$, $\Gamma_7\ (=\Gamma^7)$, is defined by:
\be
\Gamma_7 = \Gamma_0\Gamma_1\Gamma_2\Gamma_3\Gamma_5\Gamma_6 = 
\left(\begin{array}{cc}
\gamma^5 & 0 \\
0 & -\gamma^5 
\end{array}\right)\ .
\ee
\vspace{0.1cm}


\subsection*{Localization of zero modes}

Starting from eqs. (\ref{fstorder}), we obtain the following 
second order partial differential equations, holding in the whole
$(x_5,x_6)$ plane:
\bea
(\partial_5^2 + \partial_6^2) \ \chi_L^{\alpha} 
- |\mu_{(\alpha)}|^2 \ \epsilon^2(x_6) \ \chi_L^{\alpha}
- 2 \ i \ \mu_{(\alpha)} \ (-1)^k \ \delta_k(x_6) \ \eta_L^{\alpha} &=& 0 
\nonumber \\
(\partial_5^2 + \partial_6^2) \ \eta_L^{\alpha} 
- |\mu_{(\alpha)}|^2 \ \epsilon^2(x_6) \ \eta_L^{\alpha}
+ 2 \ i \ \mu_{(\alpha)}^* \ (-1)^k \ \delta_k(x_6) \ \chi_L^{\alpha} &=& 0 
\nonumber \\
(\partial_5^2 + \partial_6^2) \ \chi_R^{\alpha} 
- |\mu_{(\alpha)}|^2 \ \epsilon^2(x_6) \ \chi_R^{\alpha}
+ 2 \ i \ \mu_{(\alpha)}^* \ (-1)^k \ \delta_k(x_6) \ \eta_R^{\alpha} &=& 0 
\nonumber \\
(\partial_5^2 + \partial_6^2) \ \eta_R^{\alpha} 
- |\mu_{(\alpha)}|^2 \ \epsilon^2(x_6) \ \eta_R^{\alpha}
- 2 \ i \ \mu_{(\alpha)} \ (-1)^k \ \delta_k(x_6) \ \chi_R^{\alpha} &=& 0 
\label{app-eom}
\eea
where $k$ is an integer, $\delta_k(x_6) \equiv \delta(x_6 - k \pi R_6)$ and
the sum over $k$ is understood. In the bulk these equations are
decoupled and identical for all fields. Away from the lines $x_6 = k \pi R_6$
$k\in Z$, they read:
\be
(\partial_5^2 + \partial_6^2) \ \phi
- |\mu_{(\alpha)}|^2 \ \phi = 0
\label{app-eombulk}
\ee
with appropriate boundary conditions. Here $\phi$ stands for
$\chi_L^{\alpha}$, $\eta_L^{\alpha}$, $\chi_R^{\alpha}$,
$\eta_R^{\alpha}$.  
In each strip $k \pi R_6 < x_6 < (k+1) \pi R_6$, the
general solution to this equation can be
written in the form:
\be
\phi^{(k)}(x,x_5,x_6) = \sum_{n\in Z} \
\left[ \ C^{(k)}_n(x) \ e^{\dd\alpha_n \frac{x_6}{R_5}} +
  C'^{(k)}_n(x) \ e^{\dd-\alpha_n \frac{x_6}{R_5}}\ \right] \
e^{\dd i n \frac{x_5}{R_5}}
\label{app-sol}
\ee
with $\alpha_n = \sqrt{n^2 + |\mu_{\alpha}|^2 R_5^2}$. 
These solutions can be glued together by imposing
periodicity along $x_6$, $Z_2$ parity
and the appropriate discontinuity across the lines 
$x_6 = k \pi R_6$. This last requirement can be directly derived
from eqs. (\ref{fstorder}) and eqs. (\ref{app-eom}).
The fields $\phi$ should be continuous everywhere,
whereas their first derivatives have discontinuities
$\Delta^{(k)}(\partial_6\phi)$ given by: 
\be
\begin{array}{ll}
\Delta^{(2k)}(\partial_6\phi) = -2\, i\, s(\phi ')\, \phi '(2k\pi R_6) \quad &
 \textrm{at}\quad x_6=2k\pi R_6 \\
\Delta^{(2k+1)}(\partial_6\phi) = 2\, i\, s(\phi ')\, \phi '((2k+1)\pi R_6) \quad &
 \textrm{at}\quad x_6=(2k+1)\pi R_6
\end{array}
\label{app-jumps}
\ee
where $(\phi,\phi ') = (\chi_L^{\alpha},\eta_L^{\alpha}),\, 
                       (\eta_L^{\alpha},\chi_L^{\alpha}),\,
                       (\chi_R^{\alpha},\eta_R^{\alpha}),\, 
                       (\eta_R^{\alpha},\chi_R^{\alpha})$
and
\be
s(\phi ') = \left\{
\begin{array}{ll}
-\mu_{\alpha}\  & \textrm{if}\ \phi '=\eta_L^{\alpha},\, \chi_R^{\alpha}\\
\mu_{\alpha}^*\ & \textrm{if}\ \phi '=\eta_R^{\alpha},\, \chi_L^{\alpha}
\end{array}\right. .
\label{app-esse}
\ee
Only for $n=0$ these requirements have a non-trivial solution.
This means that the zero modes are independent of $x_5$. 
More precisely, $Z_2$-odd fields are identically
vanishing, while for even fields we get:
\be
\begin{array}{lcl}
\eta^{\alpha}_{L,R}(x,x_5,x_6) & = 
 & N^{(\alpha)}_1 f^{(\alpha)}_1(x)\, e^{\dd|\mu_{\alpha}|\, |x_6|_{per}}\\ 
&+&\, 
   N^{(\alpha)}_2 f^{(\alpha)}_2(x)\, e^{\dd-|\mu_{\alpha}|\, |x_6|_{per}} \\[0.2cm]
\chi^{\alpha}_{L,R}(x,x_5,x_6) & = 
 & i\, \dd\frac{s^*(\chi^{\alpha}_{L,R})}{|\mu_{\alpha}|}\, 
          N^{(\alpha)}_1 f^{(\alpha)}_1(x)\, e^{\dd|\mu_{\alpha}|\, |x_6|_{per}}\\ &-&\, 
   i\, \dd\frac{s^*(\chi^{\alpha}_{L,R})}{|\mu_{\alpha}|}\,
          N^{(\alpha)}_2 f^{(\alpha)}_2(x)\, e^{\dd-|\mu_{\alpha}|\, |x_6|_{per}}\, ,
\end{array}
\label{app-finalsol}
\ee
where 
$f^{(\alpha)}_{1,2}(x)$ are $x$-dependent spinors and $N^{(\alpha)}_{1,2}$
denote normalization constants, which are explicitly given in eq. (\ref{xi})
of the text.
%

%
\end{document}